\documentclass[%
twocolumn, %
aip, %
superscriptaddress, %
showpacs
]{revtex4}

\usepackage{bm,amssymb,amsmath}
\usepackage{graphicx}
\usepackage{wrapfig}
\usepackage{color}
\usepackage[dvipdfm]{hyperref}

\newcommand{\p}{\partial}

\newcommand{\di}{d_{\mathrm{i}}}
\newcommand{\de}{d_{\mathrm{e}}}
\newcommand{\rhoi}{\rho_{\mathrm{i}}}
\newcommand{\rhoe}{\rho_{\mathrm{e}}}
\newcommand{\rhos}{\rho_{\mathrm{Se}}}

\newcommand{\lcs}{\delta}

\def\Eq.#1{Eq.~(\ref{#1})}
\def\Eqs.#1#2{Eqs.~(\ref{#1}--\ref{#2})} 
\def\Eq#1{(\ref{#1})}
\def\Eqs#1#2{(\ref{#1}--\ref{#2})}

\begin{document}

\date{\today}

 \title{Gyrokinetic simulations of the tearing instability}
 \author{Ryusuke~Numata}
 \email{ryusuke.numata@gmail.com}
 \affiliation{Center for Multiscale Plasma Dynamics,
 University of Maryland,
 College Park, MD 20742, USA}
 \affiliation{Wolfgang Pauli Institute, University of Vienna, A-1090
 Vienna, Austria}
 \author{William~Dorland}
 \affiliation{Center for Multiscale Plasma Dynamics,
 University of Maryland,
 College Park, MD 20742, USA}
 \author{Gregory~G.~Howes}
 \affiliation{Department of Physics and Astronomy, University of Iowa,
 Iowa City, IA 52242, USA}
 \author{Nuno~F.~Loureiro}
 \affiliation{Associa\c{c}\~ao EURATOM/IST, Instituto de Plasmas e 
 Fus\~ao Nuclear --- Laborat\'orio Associado, Instituto Superior T\'ecnico, 
Universidade T\'ecnica de Lisboa,
1049-001 Lisboa, Portugal} 
 \author{Barrett~N.~Rogers}
 \affiliation{Department of Physics and Astronomy, Dartmouth College,
 Hanover, NH 03755, USA}
 \author{Tomoya~Tatsuno}
 \affiliation{Center for Multiscale Plasma Dynamics,
 University of Maryland,
 College Park, MD 20742, USA}

\begin{abstract}
 Linear gyrokinetic simulations covering the collisional --
 collisionless transitional regime of the 
 tearing instability are performed. It is shown that the growth rate
 scaling with collisionality agrees well with that predicted by
 a two-fluid theory for a low plasma beta case in which ion kinetic
 dynamics are negligible. Electron wave-particle interactions (Landau
 damping), finite Larmor radius, and other kinetic effects invalidate 
the fluid theory in the collisionless regime, in which a general
 non-polytropic equation of state for pressure (temperature)
 perturbations should be considered. 
We also vary the ratio of the background ion to electron temperatures, 
and show that the
scalings expected from existing calculations can be
recovered, but only in the limit of very low beta.
\end{abstract}

\maketitle

\section{Introduction}
The tearing instability~\cite{FurthKilleenRosenbluth_63} is important in magnetic 
fusion devices, where it drives the
formation of magnetic islands that can significantly degrade heat and 
particle confinement~\cite{Waelbroeck_09}. 
The related micro-tearing mode~\cite{DrakeLee_77} may lead to background 
turbulence and is also a source of confinement loss.
Solar flares and substorms in the
Earth's magnetosphere are some of the many other contexts where
tearing plays a crucial role,
inducing magnetic reconnection, explosive energy release and large-scale 
reconfiguration of the magnetic field~\cite{Biskamp_00}.

The evolution of the tearing instability critically
depends on the relationship between the width of the current layer,
$\lcs$, where the frozen-flux condition breaks down and reconnection takes place, 
and the length scales characteristic of kinetic or 
non-magnetohydrodynamic (MHD) effects, such as the ion and electron skin-depths, $\di=c/\omega_{pi}$
and $\de=c/\omega_{pe}$, the ion-sound Larmor radius, $\rho_s=c_s/\Omega_{ci}$, and the ion and
electron Larmor radii, $\rhoi$ and $\rhoe$ (see below for precise
definitions of these scales; for $T_e \sim T_i$, $\rho_s\sim \rhoi$).
For sufficiently large electron-ion collision frequency, $\nu_e$,
the width of the reconnection layer well-exceeds all of these
non-MHD scales
and the mode, at least in the limit of strong guide-field and sufficiently low $\beta$,  
is expected to be well described by
well-known resistive MHD theory~\cite{FurthKilleenRosenbluth_63,Rutherford_73,Coppi_76,Waelbroeck_89,Militello_04,Escande_04,LoureiroCowleyDorland_05}. 
In many plasmas of interest (e.g., the Earth's magnetosphere, 
modern large tokamaks), however, this is not the case:
a decrease in the collisionality of the plasma leads to a decrease in
the resistivity, causing the current layer width to shrink until it
reaches or falls below the largest relevant non-MHD scale; in the
strong guide-field case of interest here, this scale is typically 
(for $\beta>m_e/m_i$) the ion-scale
$\lambda_i$, where $\lambda_i\sim\rho_s\sim \sqrt{\beta}d_i$ 
for $\beta\ll 1$ and $\lambda_i\sim d_i$ for $\beta\sim 1$ and larger.
For plasmas in which $T_{0i}/T_{0e}\equiv\tau \gtrsim 1$
($T_{0i,e}$ are the ion and electron background temperatures),
$\lcs\lesssim\rhoi$ in this case, suggesting that a kinetic
treatment of the ions dynamics is necessary. 

Given the complexity of a fully kinetic treatment, 
a variety of simplified models have been employed to 
analytically describe tearing in such cases. 
These range from cold-ion
($e.g.$, \cite{AhedoRamos_09,Fitzpatrick_10}) 
or warm-ion ($e.g.$, \cite{LoureiroHammett_08,Sarto_11})
two-fluid approximations to calculations that include some sub-set
of ion and electron kinetic effects
(notably perpendicular ion FLR effects or electron Landau damping)
~\cite{DrakeLee_77,Cowley_86,Porcelli_91,ZakharovRogers_92,ZoccoSchekochihin_11}.
A further complication of the $\lcs\ll\lambda_i$ regime, however, is
that the tearing mode becomes coupled to pressure perturbations
related to electron and/or ion diamagnetic drifts, for example, and most existing analytic calculations
deal with this by invoking
some type of \textit{ad-hoc} closure assumption, $e.g$ isothermal or adiabatic electron
or ion equations of state. Such closures potentially play an even greater role
as $\beta$ approaches or exceeds unity, in which coupling to
the sound-waves (slow waves) along the magnetic field is also typically
important. Here we find, based on fully gyrokinetic simulations
of the linear tearing mode across a range of parameters, that when two-fluid effects
become non-negligible ($\lcs\lesssim\lambda_i$), there
is not, in general, a simple relationship between the pressure 
and density fluctuations for either the ions or electrons. 
The ratio between the two becomes a complicated function
of position that cannot be described 
by simple closure relations. This poses a serious
challenge to theoretical studies of collisionless or
weakly collisional reconnection,
particularly at higher $\beta\sim 1$ and $T_{e0}\lesssim T_{0i}$
where $\lcs\sim \rhoe$,
since a rigorous treatment would
seem to require fully kinetic treatments of both the perpendicular
and parallel electron and ion responses. 
The gyrokinetic ion and electron model used
here allows us to explore numerically, over a range of plasma parameters in
the strong guide-field limit of a simple slab geometry,
the kinetic physics of the tearing mode and the applicability of
some existing theories as the system transitions
from the collisional to collisionless regimes. \\

\section{Numerical setup}
We carry out simulations in 
doubly-periodic slab geometry using the gyrokinetic code {\tt
AstroGK}~\cite{NumataHowesTatsuno_10}. 
The equilibrium magnetic field profile is
\begin{equation}
\bm B=B_{z0}\hat z+B_{y}^{\mathrm{eq}}(x)\hat y, \quad B_{z0}\gg
 B_{y}^{\mathrm{eq}},
\end{equation}
where $B_{z0}$ is the background magnetic guide field and
$B_{y}^{\mathrm{eq}}$ is the in-plane, reconnecting component, related
to the parallel vector potential by $B_{y}^{\mathrm{eq}}(x)=-\p
A_{\parallel}^{\mathrm{eq}}/\p x$. From the formal point of view, 
$B_{y}^{\mathrm{eq}}$ is a first-order gyrokinetic perturbation. To set it up, 
we perturb the background (Maxwellian) electron distribution function $f_{0\mathrm{e}}(v)$ 
with a shifted Maxwellian $\delta f_{\mathrm{e}} \propto v_{\parallel}$
($v$, $v_{\parallel}$ are the velocity-space coordinates), yielding
 \begin{equation}
  \label{equilib_apar}
   A_{\parallel}^{\mathrm{eq}}(x) =
   A_{\parallel0}^{\mathrm{eq}}\cosh^{-2}\left(\frac{x-L_{x}/2}{a}\right)
   S_{\mathrm{h}}(x),
 \end{equation}
where $A_{\parallel0}^{\mathrm{eq}}=3\sqrt{3}/4$ (such that $\max| B_y^{\mathrm{eq}}(x)|=1.0$), and
$S_{\mathrm{h}}(x)$ is a shape function to enforce periodicity~\cite{NumataHowesTatsuno_10}.
The equilibrium scale length is denoted by $a$ and $L_x$ is the domain length  
in the $x$-direction, set to $L_x/a=3.2\pi$. In the $y$-direction,
we set $L_y/a=2.5\pi$, resulting in a value of the tearing instability parameter
$\Delta'a \approx 23.2$~\cite{LoureiroCowleyDorland_05} for the longest
wavelength mode in the system: $k_y a=2\pi a/L_y=0.8$.
Constant $background$ temperatures ($T_{0\mathrm{i},\mathrm{e}}$) and
densities ($n_{0\mathrm{i},\mathrm{e}}$) are assumed for both
species. We consider a quasi-neutral plasma, so
$n_{0{\mathrm{i}}}=n_{0{\mathrm{e}}}=n_0$, and singly charged ions
$q_{\mathrm{i}}=-q_{\mathrm{e}}=e$.

We employ a model collision operator which satisfies physical
requirements~\cite{AbelBarnesCowley_08,BarnesAbelDorland_09} and is
able to reproduce Spitzer
resistivity~\cite{SpitzerHarm_53}, for which the electron-ion collision
frequency ($\nu_{\mathrm{e}}$) and the resistivity ($\eta$) are related by
\begin{equation}
 \eta/\mu_{0} = 0.380 \nu_{\mathrm{e}} d_{\mathrm{e}}^{2}
  \label{eq:spitzer_resistivity}
\end{equation}
with $\mu_{0}$ being the vacuum permeability; this formula holds for
$k_{\perp}d_{\mathrm{e}}\lesssim1$ ($k_{\perp}^{-1}$ is a characteristic scale
length perpendicular to a mean magnetic field, $k_{\perp}^{-1}\sim\lcs$ for the 
tearing instability). Like-particle collisions are neglected.

{\tt AstroGK} employs a pseudo-spectral algorithm to discretize the GK
equation in the spatial coordinates $(x,y)$. For the linear runs
reported here, it is sufficient to keep only one harmonic in the
$y$-direction (the lowest harmonic is the fastest growing one).
The number of Fourier modes in the $x$-direction ranges from 
$512$ to $8192$ (multiplied by $2/3$ for dealising).
Velocity space integrals are evaluated using Gaussian quadrature; the
velocity grid is fixed to $20\times16$ collocation points in the
pitch-angle and energy directions, respectively. Convergence tests
have been performed in all
runs to confirm the accuracy of our results. \\

\section{Problem setup}
We scan in collisionality and use Eq.~\eqref{eq:spitzer_resistivity}
to calculate the plasma resistivity $\eta$, recast in terms of the
Lundquist number,
$S=\mu_{0}aV_{\mathrm{A}}/\eta
=2.63(\nu_{\mathrm{e}}\tau_{\mathrm{A}})^{-1}(d_{\mathrm{e}}/a)^{-2}$,
where $V_{\mathrm{A}}$ is the Alfv\'en velocity corresponding to the
peak value of $B_{y}^{\mathrm{eq}}$ and 
$\tau_{\mathrm{A}} \equiv a/V_{\mathrm{A}}$ is the Alfv\'en time.
Other relevant quantities are:
\begin{equation}
 \label{eq:ion_e_scales}
\begin{split}
 \rhoi = \tau^{1/2} \rhos\sqrt{2}, ~~~ &
 d_{\mathrm{i}} = \beta_{\mathrm{e}}^{-1/2} \rhos\sqrt{2}, \\
 \rhoe = \sigma^{1/2} \rhos\sqrt{2}, ~~~ &
 d_{\mathrm{e}} = \beta_{\mathrm{e}}^{-1/2} \sigma^{1/2} \rhos\sqrt{2},
\end{split}
\end{equation}
where $\rho_{i,e}$ and $d_{i,e}$ are the ion and electron Larmor radii and 
skin-depths, respectively,
$\sigma \equiv m_{\mathrm{e}}/m_{\mathrm{i}}$,  
$\tau \equiv T_{0\mathrm{i}}/T_{0\mathrm{e}}$, 
$\beta_{\mathrm{e}}\equiv
n_{0}T_{0\mathrm{e}}/(B_{z0}^{2}/2\mu_{0})$, $\rhos\equiv
c_{\mathrm{Se}}/\Omega_{\mathrm{ci}}$, 
$c_{\mathrm{Se}}=\sqrt{T_{0\mathrm{e}}/m_{\mathrm i}}$, 
$\Omega_{\mathrm{ci}}=e B_{z0}/m_{\rm i}$, $\rho_s=\rhos\sqrt(1+\tau)$. 
In addition to $\nu_{\mathrm{e}}$, the adjustable parameters
considered here include the mass ratio $\sigma$, the electron
beta $\beta_{\mathrm{e}}$, $\rhos/a$, and $\tau$, although the latter is
held fixed at $\tau=1$ except where stated otherwise.

We study the collisional--collisionless transition by scanning in collisionality.
As $\nu_e$ is decreased, the different ion and electron kinetic scales become important.
Given the challenge of clearly separating all the relevant spatial 
scales in a kinetic simulation,
we split our study into two sets of runs: a smaller-$\rhos$ series
($\rhos/a=0.02/\sqrt{2}\simeq 0.014$) and a larger-$\rhos$ series ($\rhos/a=0.2/ \sqrt{2}\simeq 0.14$).
Since $\tau=1$ is held fixed except at the end of the article,
these two sets of runs also typically correspond to $\rhoi/a=0.02$ and $\rhoi/a=0.2$, 
respectively.

In the former set $\rho_e,~d_e\ll \lambda_{\mathrm{i}} \lesssim \lcs \ll a$;
in this case the frozen-flux condition is broken by collisions alone, and
since $\lcs$ well exceeds the collisionless electron
scales $\rho_e,~d_e$, such scales need
not be resolved in the simulations. The ion response, on the other hand,
is predominantly collisional ($\lcs >\lambda_i$) at the smallest
considered values of $S\sim 500$ but kinetic
($\lcs\lesssim \lambda_i$) at the largest values, $S\sim 10^5$.
Thus resistive MHD would be expected, at least marginally,
to be valid in this case at the smaller $S$ values. 
In the set of runs with larger-$\rhos$ ($\rhos/a\simeq 0.14$), we again consider
$\rho_e,~d_{\mathrm{e}}\ll\lambda_{\mathrm{i}} \lesssim a$, but since
$\rhos/a$ is ten times larger than in the previous set of runs,
the ions in this second set are predominantly kinetic 
($\lcs\lesssim \lambda_i$) over the entire
considered range of $S\sim 100-10^6$. Indeed, at the highest values of $S$,
$\lcs$ reaches collisionless electron scales ($d_e$ at $\beta\ll 1$ and
$\rho_e$ at $\beta\sim 1$), and the
instability dynamics become essentially collisionless.

In both sets of runs, we vary $S$ over the ranges mentioned above
for three different sets of
$\beta_{\mathrm{e}}$ and $\sigma=m_e/m_i$:  [($\beta_{\mathrm{e}}$,
$\sigma$)=(0.3,0.01), (0.075,0.0025), (0.01875,0.000625)].
These parameters are such that 
$\rhos/d_e \equiv \sqrt{\beta_e/(2\sigma)}=\sqrt{15}\simeq 3.9$
is held fixed and thus, since $\rho_s/a$ is also held fixed
(at either 0.014 or 0.14), $d_e/a$ is also held fixed
(at either 0.0037 or 0.037, respectively). Given the parameter dependences
of $d_i$ and $\rho_e$ noted in Eq.~\ref{eq:ion_e_scales}, however, it is seen
that the values of $d_i/a$ and $\rho_e/a$ both change as $\beta_e$
and $\sigma$ are varied in this manner: for $\rhos=0.014$, $d_i=0.02/\sqrt{\beta_e}$
and $\rho_e/a=0.02\sqrt{\sigma}$, while for $\rhos/a=0.14$ they are ten times
larger. \\

\section{Smaller $\rhos/a=0.014$}
\begin{figure*}[t!]
 \begin{center}
  \includegraphics[scale=0.7]{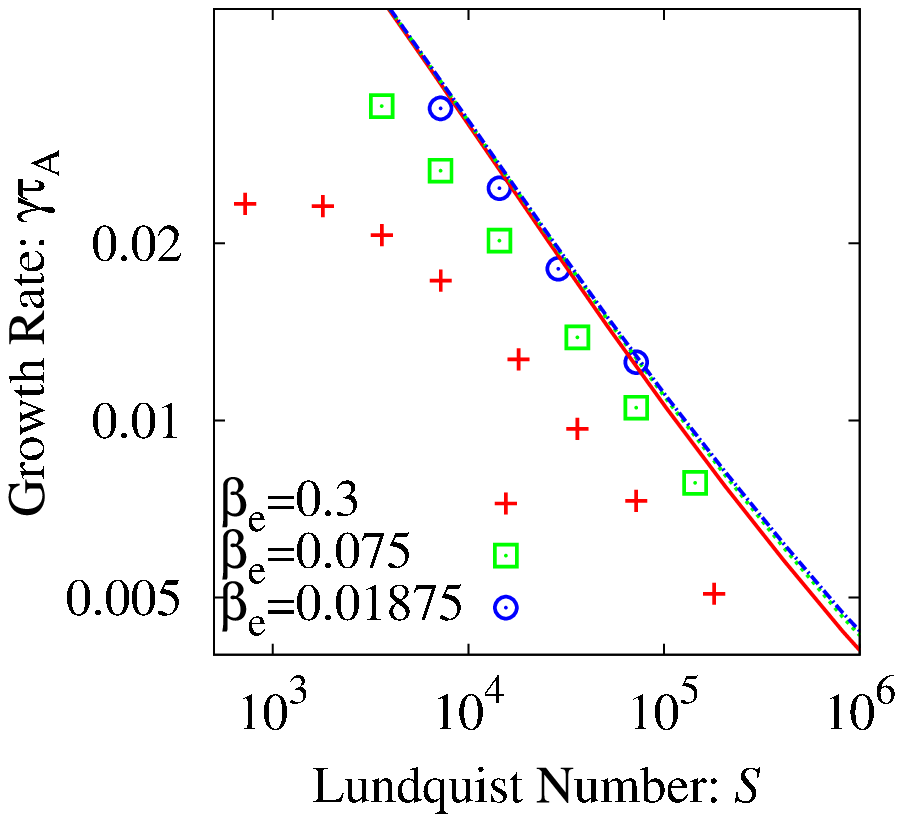}
  \hskip 0.4cm
  \includegraphics[scale=0.7]{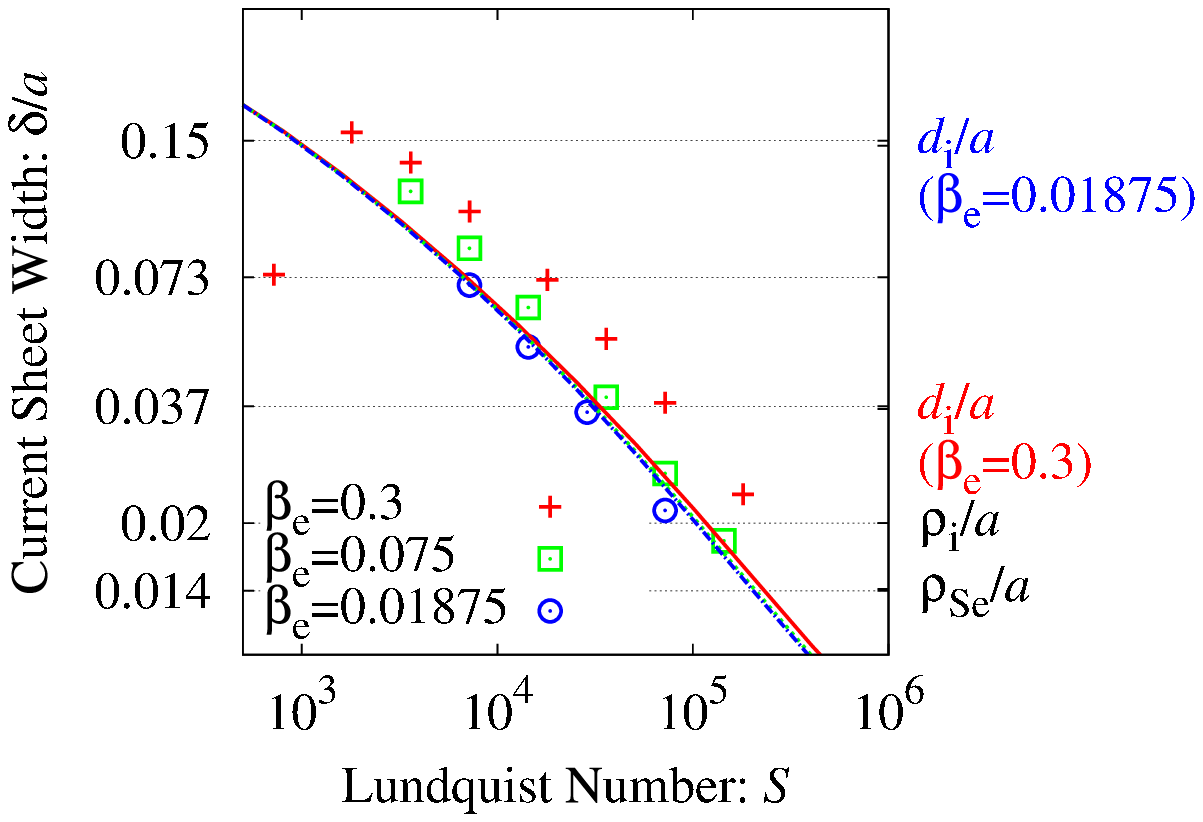}
  \caption{\label{fig:growth_layer_r0.02}(Color online)
  Growth rate and current sheet width versus the Lundquist number for
  $\rhos/a=0.02/\sqrt{2}$. Red crosses, green squares, and blue circles
  show gyrokinetic results for $(\beta_{\mathrm{e}},\sigma)=(0.3,0.01),
  (0.075,0.0025), (0.01875, 0.000625)$, respectively. Red solid, green
  dashed and blue dot-dashed lines are the corresponding two-fluid~\cite{Fitzpatrick_10} 
  scalings.
  The relevant  scale lengths are  identified on the right axis of the right panel.
}
 \end{center}
\end{figure*}
Fig.~\ref{fig:growth_layer_r0.02} shows the tearing mode growth rate 
($\gamma\tau_A=d \log A_{\parallel}/dx$ evaluated at the $X$-point)
and current layer width (full-width at half-maximum - typically somewhat larger
than the current-profile scale-length, depending on the current profile)
as functions of the Lundquist number (symbols) for $T_{i0}=T_{e0}$.
Also plotted (lines) are the results obtained from a reduced
two fluid model~\cite{Fitzpatrick_10} with an isothermal electron equation of state. 
This model is derived under the assumption
of low-$\beta$, but exactly how low $\beta$ must be for the validity
of this model depends on how the various quantities in the
model are ordered and is thus problem-dependent.
For the ordering assumed in~\cite{Fitzpatrick_10},
it is argued that
$\beta_e\ll \sqrt{\sigma}$ is required --- a condition that is marginally
satisfied here only for the lowest $\beta$ case,
$(\beta_{\mathrm{e}},\sigma)=(0.01875, 0.000625)$.
The two fluid model is also derived under the assumption of
cold ions, but as we show later (see fig.~\ref{fig:ti_dep}), the difference between
the gyrokinetic results at $\tau=0$ and $\tau=1$ 
is small. 

For the largest value of the collisionality, $S\lesssim 10^3$,
the gyrokinetic growth rates roll-over because 
the current layer width is too wide to satisfy the asymptotic scale
separation, $\lcs\ll a$, assumed in the two-fluid model tearing mode dispersion relation
that is plotted in the figure.
The deviation between the gyrokinetic and two-fluid results at the lowest $S$ values
should therefore be disregarded. It is seen from the right panel that, as noted earlier,
$\lcs>\lambda_i$ for all but the largest $S$ values
(recall that $\lambda_i$, the outer-most collisionless ion-scale of relevance,
is typically $\lambda_i\sim\rhos\sim \rhoi \ll d_i$ for $\beta\ll 1$ and 
$\lambda_i\sim \rhos\sim d_i$ for $\beta_e=0.3$).
In this case, as expected, the two-fluid model, at least at low-$\beta$, recovers the well-known
single-fluid resistive-MHD scalings\cite{FurthKilleenRosenbluth_63} and is thus
independent of $\beta_{\mathrm{e}}$. Defining the dimensionless parameter
$C_k=k_y a^2 B'(0)/B_0$ where $B'(0)$ is the derivative of
the equilibrium reconnecting field at the X-point, $k_y$ is the linear mode-number,
and $B_0$ is the normalizing field (for our parameters $k_y a=0.8$ and
and $aB'(0)/B_0=2.6$ so that $C_k=2.08$ 
for all runs), the general one-fluid scalings are most
compactly written in terms of the quantities $\tau_{Ak}=a/(V_A C_k)=\tau_A/C_k$
and $S_k=C_k S$. Two scalings are obtained depending 
on the product of $\Delta'\delta$; defining $C_\Delta=\Delta' a/2$ (equal to 11.6 in our runs), 
they are
\begin{align}
&\gamma\tau_{Ak}\simeq S_k^{-1/3}\ ,~ \delta/a\sim S_k^{-1/3} \text{ for } \Delta'\lcs/2 \gg 1,\\
\nonumber\\
&\gamma\tau_{Ak}\simeq 0.96 S_k^{-3/5}C_\Delta^{4/5}\ ,
~ \delta/a\sim S_k^{-2/5} C_\Delta^{1/5},\nonumber\\
&\qquad\qquad\qquad\qquad\qquad\qquad\qquad  \text{ for } \Delta'\lcs/2 \ll 1. 
\end{align}
For the parameters of our simulations ($C_k=2.08$ and $C_\Delta=11.6$)
these can also be written in terms of $\tau_A$ and $S$ as
\begin{equation}
\gamma\tau_{A}\simeq 1.63 S^{-1/3}\ ,~ \delta/a\sim 0.78 S^{-1/3} \text{ for } \Delta'\lcs/2 \gg 1,
\end{equation}
\begin{equation}
\gamma\tau_{A}\simeq 9.14 S^{-3/5}\ ,
~ \delta/a\sim 1.22 S^{-2/5} \text{ for } \Delta'\lcs/2 \ll 1.
\end{equation}
The small and large  $\Delta'$ expressions for $\gamma$ and $\delta/a$ both
break-down at the point of maximum growth rate, where they are
roughly equal: $S_k\sim C_\Delta^3$, or for our parameters,
$S\sim 600$. Since $S$ is larger or comparable to this value
in the simulations presented here, most of our runs are in the either
the marginally-large or small-$\Delta'$ regimes.
Using for example the small-$\Delta'$ expressions for the value $S=10,000$, we obtain
$\delta/a\simeq 0.03$ and $\gamma\tau_A \simeq 0.04$, in rough
agreement with the numerical results. The over-estimation
of the growth rates by the two fluid model at higher $\beta_e\sim 0.3$
is possibly due to either a breakdown in the low-$\beta$
ordering of the fluid model or a gradual onset of kinetic effects
($e.g.$ the invalidity of a simple isothermal equation of state,
as discussed further below).

\section{Larger $\rhos/a=0.14$}
\begin{figure*}
 \begin{center}
  \includegraphics[scale=0.7]{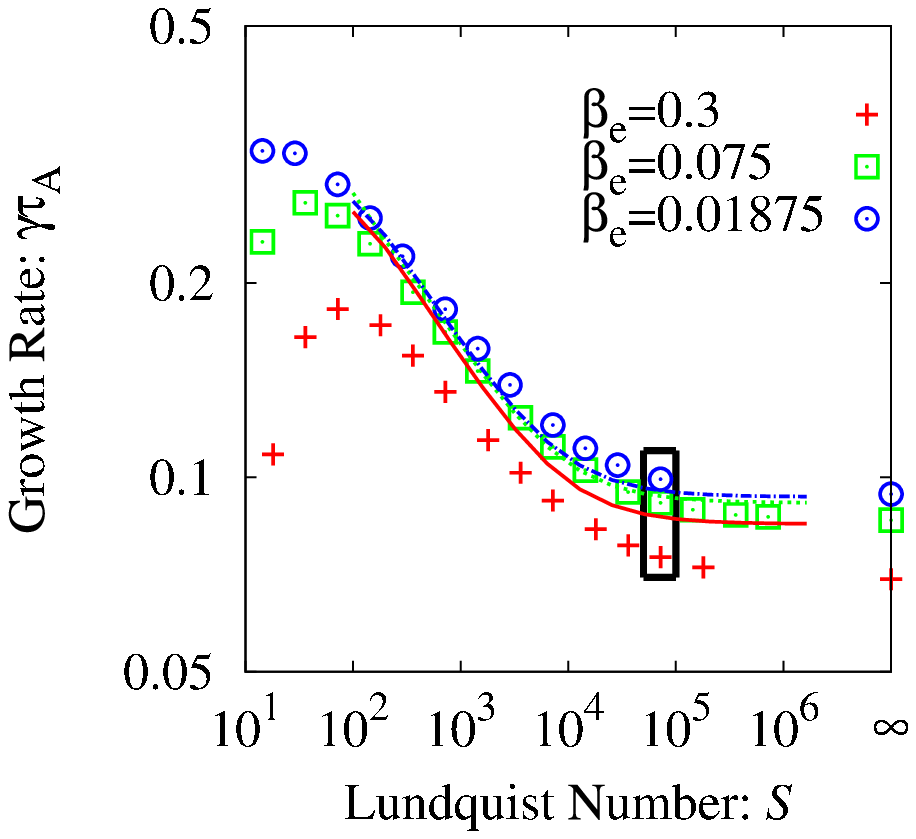}
  \hskip 0.4cm
  \includegraphics[scale=0.7]{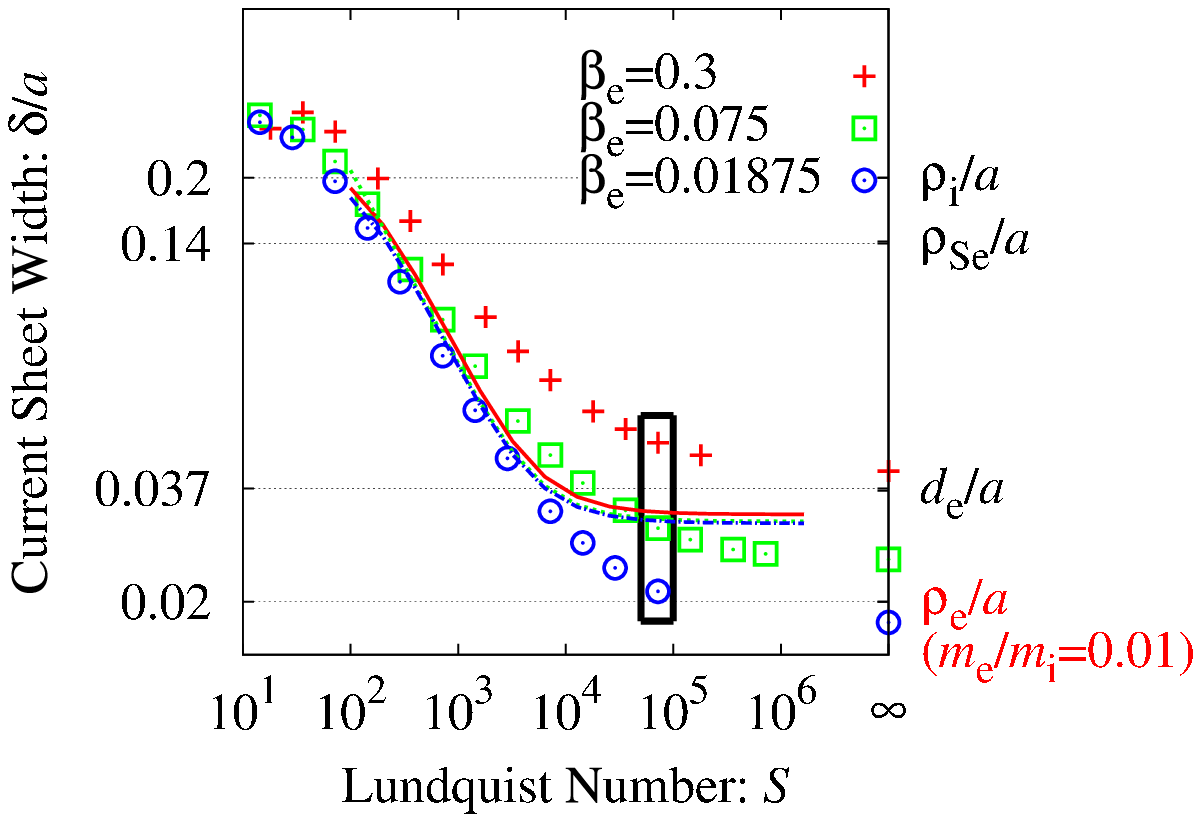}
  \caption{\label{fig:growth_layer_r0.2}(Color online)
  Growth rate and current sheet width versus the Lundquist
  number for $\rhos/a=0.2/\sqrt{2}$. Color and symbol schemes are the same
  as in Fig.~\ref{fig:growth_layer_r0.02}. The scale lengths
  labelled on the right panel are fixed except $\rho_{\mathrm{e}}$ 
 (only $\rho_{\mathrm{e}}$ for $\sigma=0.01$ is
  shown). The points in the box are used to diagnose temperature
  fluctuations in Fig.~\ref{fig:polytropic_indices}.}
 \end{center}
\end{figure*}

We set $\rhos/a=0.2/\sqrt{2}=0.14$ and adjust $\nu_{\mathrm{e}}$ such that
$\lcs\lesssim\lambda_{\mathrm{i}}$, thus focusing on the regime where ion
kinetic effects are important. 
The growth rate and 
current layer width versus the Lundquist number are shown in Figure~\ref{fig:growth_layer_r0.2}
(the label $S=\infty$ identifies the case $\nu_{\mathrm{e}}=0$;
we note that $S$ may be underestimated if $\delta \lesssim d_{\mathrm{e}}$
since Eq.~\eqref{eq:spitzer_resistivity} is not valid in such a
regime). These runs correspond to the same set of ($\beta_{\mathrm{e}}$,
$\sigma$) as before; in terms of length scales,
$d_{\mathrm{e}}/a\approx0.037$ is fixed, and $\di$ and $\rhoe$ change.

As in the previous case, we observe better agreement between the GK and two fluid
results for lower values of $\beta_{\mathrm{e}}$; however, as $S$ increases and the 
collisionless regime is approached, the
agreement becomes poorer for any value of $\beta_{\mathrm{e}}$.
In this regime, electron
kinetic effects (Landau damping and even finite electron orbits: 
note that for $\beta_{\mathrm{e}}=0.3$, $\lcs/\rhoe\approx 2$) play an important role;
these are absent in the two fluid model.

\section{Temperature fluctuations}
\begin{figure*}
 \includegraphics[scale=0.7]{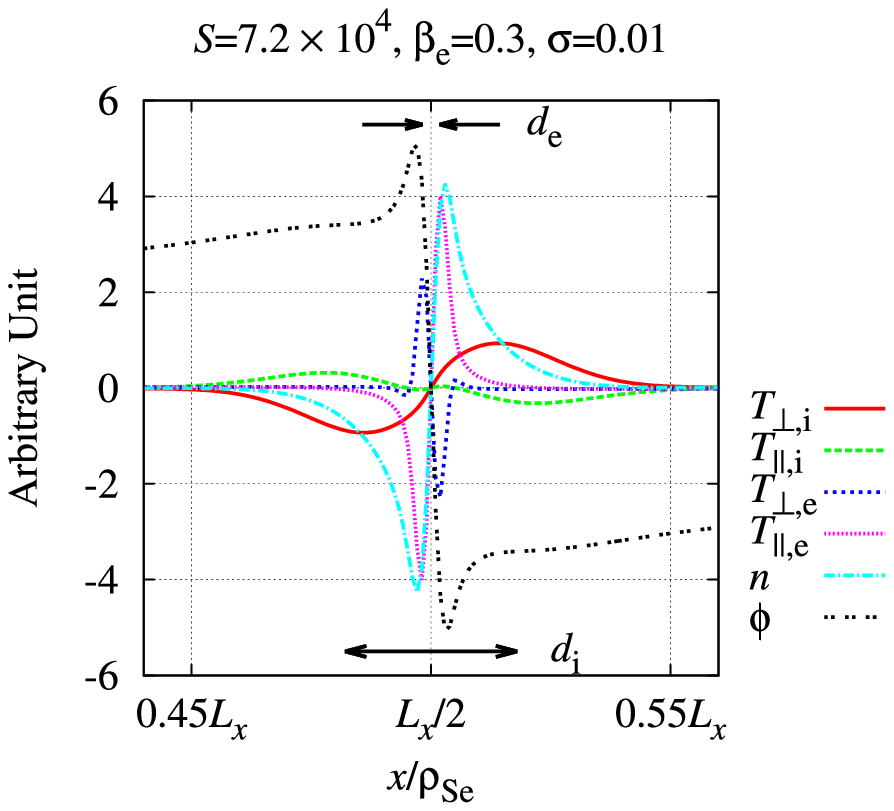}
\hskip 0.4cm
 \includegraphics[scale=0.7]{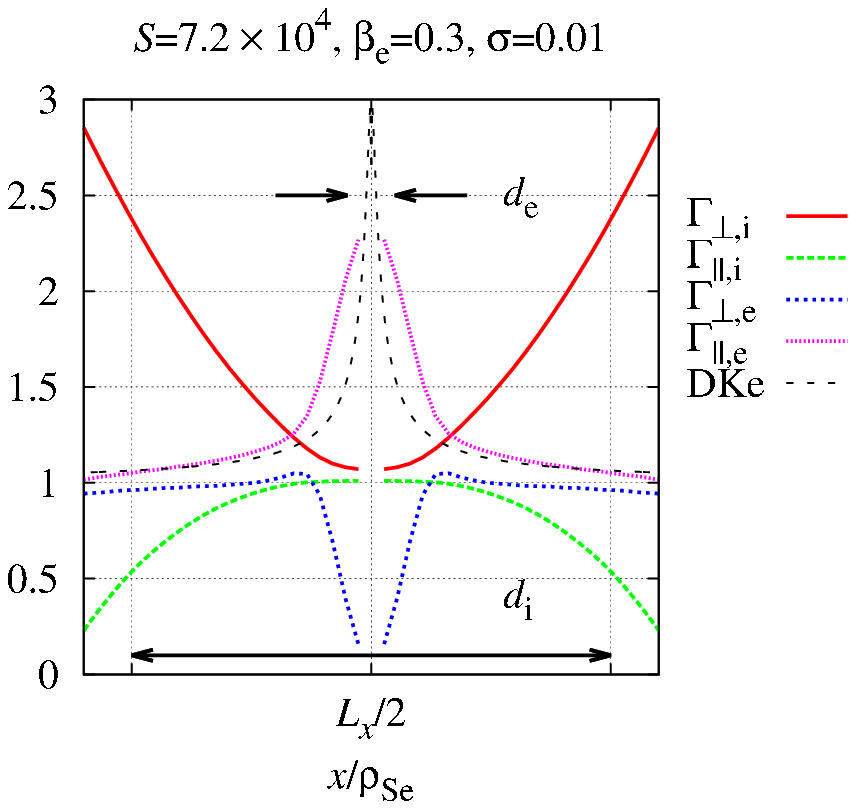}
 \includegraphics[scale=0.7]{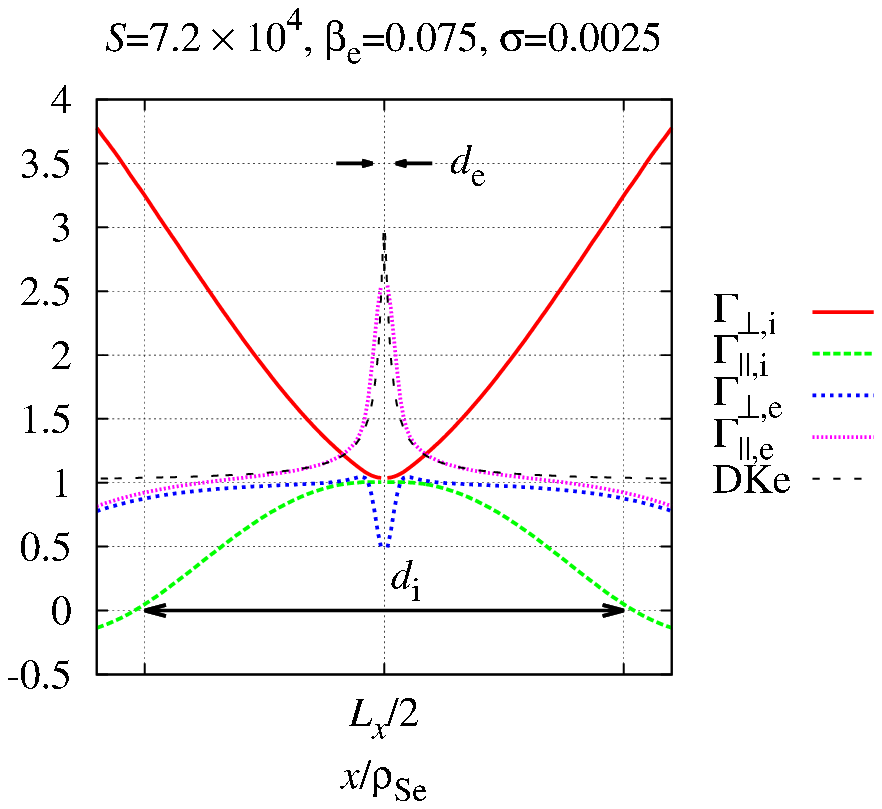}
\hskip 0.4cm
 \includegraphics[scale=0.7]{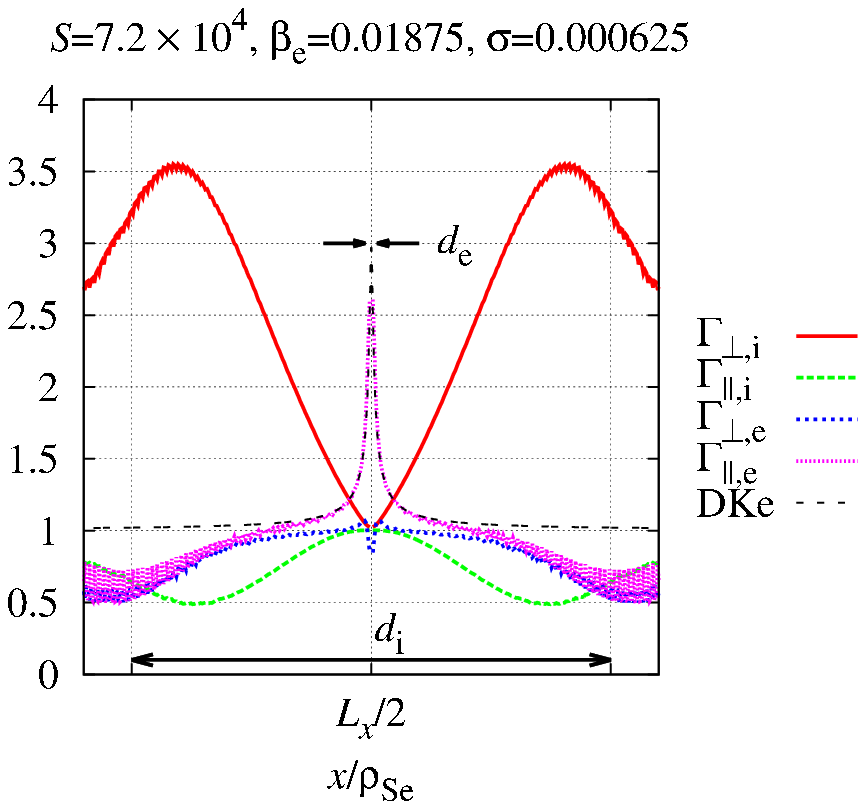}
 \caption{\label{fig:polytropic_indices}(Color online)
 Eigenfunctions for the $\beta_{\mathrm{e}}=0.3$ case,
 and the
 polytropic indices of ions and electrons in the perpendicular and the
 parallel directions.
A spatially changing polytropic index means
 the equation of state is not polytropic.}
\end{figure*}

To better understand the discrepancies between the GK and two fluid results
(at high-$\beta_{\mathrm{e}}$ regardless of the 
collisionality and at any $\beta_{\mathrm{e}}$ in the collisionless regime), 
we examine in Fig.~\ref{fig:polytropic_indices} the validity 
of the isothermal closure by diagnosing the 
temperature fluctuations for the $\rhos/a=0.2/\sqrt{2}$ and
$S\approx7.2\times10^4$ case.
In Fig.~\ref{fig:polytropic_indices}, we plot the eigenfunctions
of the temperature, density and electrostatic potential ($\phi$), and
the diagonal components of the pressure tensor fluctuations normalized by
$\tilde{n}$. These are measures of the polytropic indices
$\Gamma_{s}$ because, according to the polytropic law,
$\tilde{p}_{s}=\Gamma_{s}T_{0s}\tilde{n}_{s}$ ($s=\mathrm{i},\mathrm{e}$
is a species label). We define $\Gamma_{\perp,s}=(\Gamma_{xx,s}+\Gamma_{yy,s})/2$
and $\Gamma_{\parallel,s}=\Gamma_{zz,s}$, but restrict the discussion to 
the parallel temperature for electrons as the
perpendicular component is at least marginally smaller (by a factor
of ${\mathcal O}[(k_{\perp}\rho_{\mathrm{e}})^{2}]$) in Ohm's
law.

As seen in the figure, while $\Gamma_{\parallel,\mathrm{e}}\sim 1$ outside the layer (thus 
validating the isothermal electron approximation in that region), it is
highly peaked in the current layer
($x\sim d_{\mathrm{e}}$)  due to the Landau damping. A spatially varying 
polytropic index means that the equation of state is not polytropic, and thus
the isothermal closure is violated.
Also plotted is $\Gamma_{\parallel,\mathrm{e}}$ obtained analytically from
the drift-kinetic electron (DKe) model~\cite{DrakeLee_77}. We observe that this 
expression provides an excellent fit to our data for the case $\beta_{\mathrm{e}}=0.01875$,
where electron FLR effects are negligible.
As for the ions,  $\Gamma_{\mathrm{i}}$ also varies widely over the 
ion inertial scale in all cases, again invalidating simple equations of 
state for this species.

\section{Ion background temperature}
Finally, we examine the effect of the ion background temperature ($\tau$), 
as this is the other possible source of discrepancy between the GK and the two fluid 
results.

\begin{figure*}
 \includegraphics[scale=0.7]{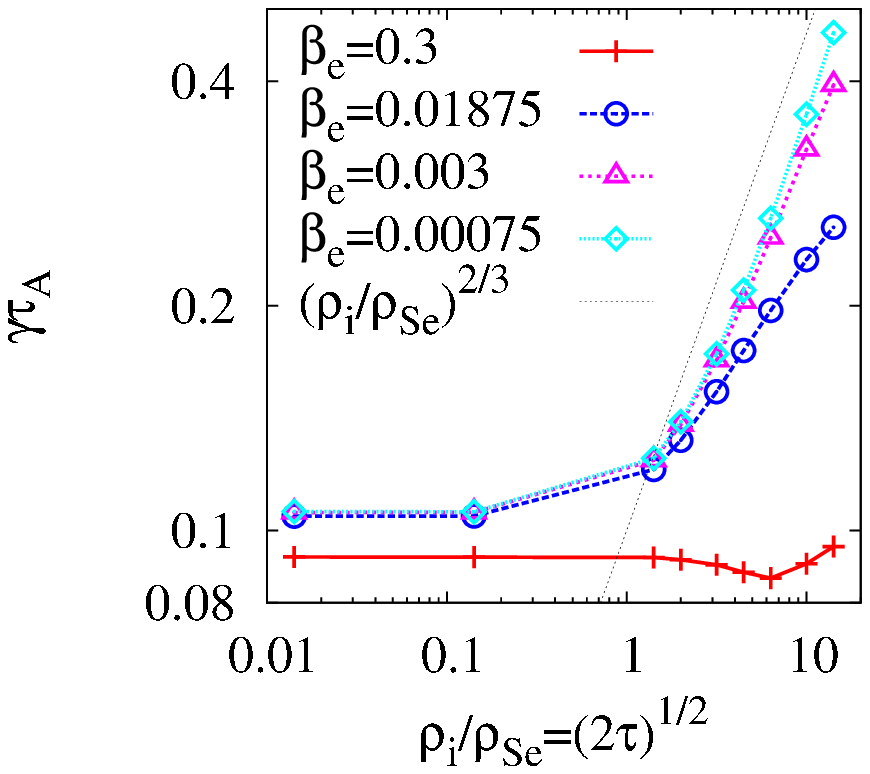}
\hskip 0.4cm
 \includegraphics[scale=0.7]{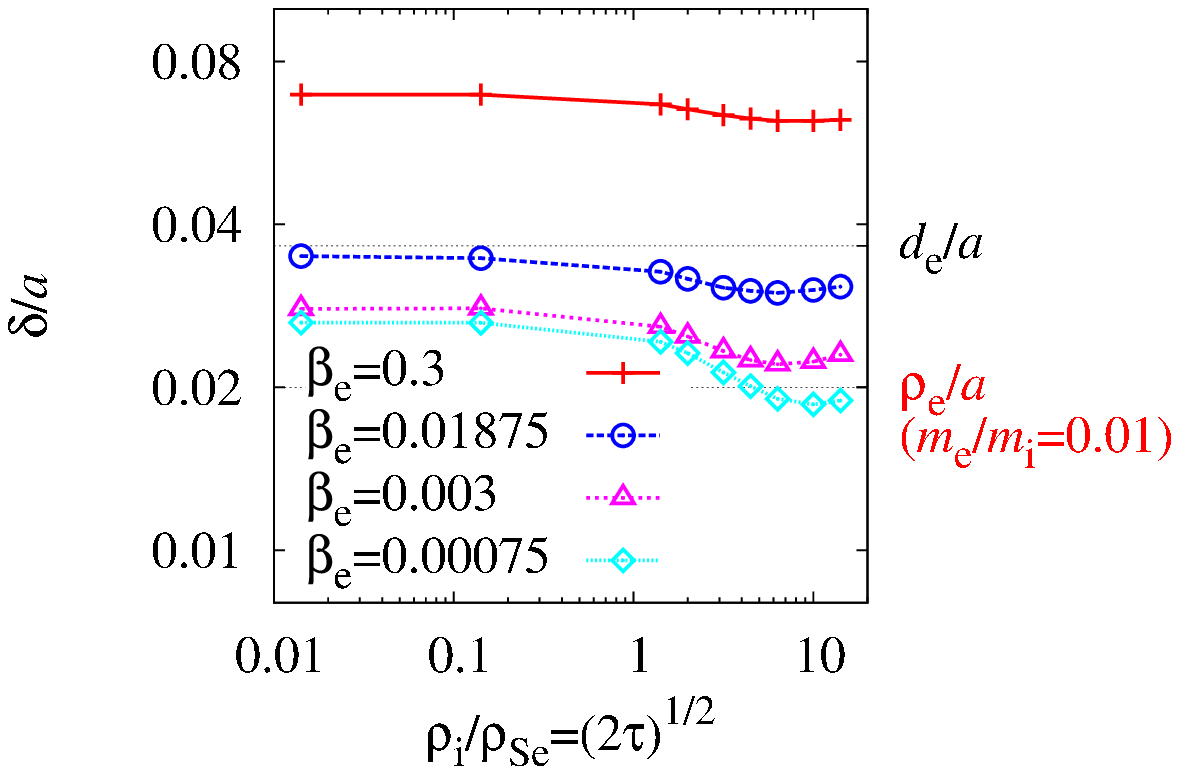}
 \caption{\label{fig:ti_dep}(Color online)
 Ion temperature dependence of the tearing growth rate and current layer
 width. 
 }
\end{figure*}

Plotted in Figure~\ref{fig:ti_dep} are the growth rate and current
layer width versus $\rho_{\mathrm{i}}/\rho_{\mathrm{Se}}$. As before, we probe different values of 
$\beta_{\mathrm{e}}$ and change $\sigma$ such that 
$\rho_{\mathrm{Se}}/a\approx0.14$, $d_{\mathrm{e}}/a\approx0.037$ in all cases. The Lundquist number is
$S\approx7200$. 
As in Ref.~\cite{RogersKobayashiRicci_07}, we find that the growth rate is remarkably 
insensitive to $\tau$ for large $\beta_{\mathrm{e}}$; 
however, as $\beta_e$ decreases we 
are able to recover the theoretically predicted
scaling~\cite{Porcelli_91} of $\gamma\tau_A\sim \tau^{1/3}$. 
One possible explanation of these results is that, at higher $\beta_e$,
coupling to sound-waves becomes more important and even more strongly
invalidates the simple closure relations used in the
fluid calculations.

\section{Discussion and Conclusions}

We have performed a set of gyrokinetic simulations of the 
linear tearing instability, varying the collisionality, plasma $\beta$,
mass ratio  $m_e/m_i$, and $T_i/T_e$.
Although we find agreement with a two-fluid description~\cite{Fitzpatrick_10} 
in the collisional low-$\beta$ case, 
one of our main conclusions is that if the plasma parameters are such that 
two fluid effects are important,
there is not, in general, a simple relationship between the pressure and 
density fluctuations, their ratio 
being a function of position. 
This is an intrinsically kinetic effect which cannot 
be captured by any known fluid closure.
It seems likely that in our simulations,
at non-small values of $\beta$, the parallel
sound wave dynamics become important and
ion flow and temperature must be solved for kinetically. In addition, 
we find that 
electron Landau damping cannot be neglected in the collisionless regime. 
If $\delta>\rho_{\mathrm{e}}$, the effects of finite electron orbits can be neglected, and the 
Landau damping effect is analytically
tractable using the drift-kinetic model (DKe)~\cite{DrakeLee_77}.
For $\delta\sim\rho_{\mathrm{e}}$ the DKe model is not sufficient and a fully kinetic treatment
is required.

We have also shown that for $\beta_{\mathrm{e}}\ll1$ (such that the ion sound wave 
and the Alfv\'en wave are decoupled), the theoretically predicted dependence 
of the growth rate on $\tau$, $\gamma\sim\tau^{1/3}$~\cite{Porcelli_91} is verified.
However, for $\beta_{\mathrm{e}}\sim 1$ the growth rate in our system is remarkably 
insensitive to the background ion temperature, as noted in  previous numerical studies
~\cite{RogersKobayashiRicci_07}.

As is widely known, linear theory breaks down
when the magnetic island width grows beyond the width
of the current layer, and indeed, in some
physical systems, turbulent noise seems likely
to generate seed magnetic islands that are
nonlinear from birth. It is therefore important
to understand the importance of kinetic effects
in the nonlinear phase --- a topic not studied
here. The answer to this, particularly in the
strong guide-field limit in which 3D kinetic
simulations are only just starting to be explored,
will likely depend on the aspect of reconnection
that is of interest. As in the case of relatively
small systems without a guide field, it may be
that gross features of the reconnection, such
as the reconnection rate, can be, at least qualitatively,
captured by fluid models. On the other hand,
in weakly collisional plasmas, as suggested by
the results found here, it seems likely that questions involving
particle heating or energy partition, for example,
will likely require a kinetic physics model that
includes effects such as Landau damping and
goes beyond simple closure schemes. The potential
strength of the gyrokinetic model --- and weakness,
in some physical problems --- is that it is designed to
study the strong guide-field limit, in which time-resolution
of the electron gyroperiod (necessary in particle
simulations, for example, but not in gyrokinetics)
can become a formidable challenge. Further work
is needed to explore nonlinear reconnection
in the strong guide-field limit, and understand the
contributions that gyrokinetic simulations may offer.
\section{Acknowledgments}
The authors thank M. Furukawa, F. Jenko, V.~V. Mirnov,
M. P\"uschel, J.~J. Ramos, Z. Yoshida, and A. Zocco for useful comments, 
and A.A. Schekochihin for numerous suggestions that much improved this manuscript.
This work was supported by the DOE Center for Multiscale Plasma Dynamics, 
the DOE-Espcor grant
to CICART,
the Leverhulme Trust Network for Magnetized Turbulence,
the Wolfgang Pauli Institute (Vienna, Austria)
Funda\c{c}\~ao para a Ci\^encia e a Tecnologia (Portugal) 
and the European Community under the contract of Association between
EURATOM and IST.
The views and opinions expressed herein do not necessarily reflect those
of the European Commission.
Simulations were performed on TACC, NICS, NCCS and NERSC supercomputers.

\providecommand{\noopsort}[1]{}


\begin{thebibliography}{23}
\expandafter\ifx\csname natexlab\endcsname\relax\def\natexlab#1{#1}\fi
\expandafter\ifx\csname bibnamefont\endcsname\relax
  \def\bibnamefont#1{#1}\fi
\expandafter\ifx\csname bibfnamefont\endcsname\relax
  \def\bibfnamefont#1{#1}\fi
\expandafter\ifx\csname citenamefont\endcsname\relax
  \def\citenamefont#1{#1}\fi
\expandafter\ifx\csname url\endcsname\relax
  \def\url#1{\texttt{#1}}\fi
\expandafter\ifx\csname urlprefix\endcsname\relax\def\urlprefix{URL }\fi
\providecommand{\bibinfo}[2]{#2}
\providecommand{\eprint}[2][]{\url{#2}}

\bibitem[{\citenamefont{Furth et~al.}(1963)\citenamefont{Furth, Killeen, and
  Rosenbluth}}]{FurthKilleenRosenbluth_63}
\bibinfo{author}{\bibfnamefont{H.~P.} \bibnamefont{Furth}},
  \bibinfo{author}{\bibfnamefont{J.}~\bibnamefont{Killeen}}, \bibnamefont{and}
  \bibinfo{author}{\bibfnamefont{M.~N.} \bibnamefont{Rosenbluth}},
  \bibinfo{journal}{Phys. Fluids} \textbf{\bibinfo{volume}{6}},
  \bibinfo{pages}{459} (\bibinfo{year}{1963}).

\bibitem[{\citenamefont{Waelbroeck}(2009)}]{Waelbroeck_09}
\bibinfo{author}{\bibfnamefont{F.~L.} \bibnamefont{Waelbroeck}},
  \bibinfo{journal}{Nucl. Fusion} \textbf{\bibinfo{volume}{49}},
  \bibinfo{pages}{104025} (\bibinfo{year}{2009}).

\bibitem[{\citenamefont{Drake and Lee}(1977)}]{DrakeLee_77}
\bibinfo{author}{\bibfnamefont{J.~F.} \bibnamefont{Drake}} \bibnamefont{and}
  \bibinfo{author}{\bibfnamefont{Y.~C.} \bibnamefont{Lee}},
  \bibinfo{journal}{Phys. Fluids} \textbf{\bibinfo{volume}{20}},
  \bibinfo{pages}{1341} (\bibinfo{year}{1977}).

\bibitem[{\citenamefont{Biskamp}(2000)}]{Biskamp_00}
\bibinfo{author}{\bibfnamefont{D.}~\bibnamefont{Biskamp}},
  \emph{\bibinfo{title}{Magnetic Reconnection in Plasmas}}
  (\bibinfo{publisher}{Cambridge Univ. Press}, \bibinfo{address}{Cambridge},
  \bibinfo{year}{2000}), ISBN \bibinfo{isbn}{0-521-58288-1}.

\bibitem[{\citenamefont{Rutherford}(1973)}]{Rutherford_73}
\bibinfo{author}{\bibfnamefont{P.}~\bibnamefont{Rutherford}},
  \bibinfo{journal}{Phys. Fluids} \textbf{\bibinfo{volume}{16}},
  \bibinfo{pages}{1903} (\bibinfo{year}{1973}).

\bibitem[{\citenamefont{Coppi et~al.}(1976)\citenamefont{Coppi, Pellat,
  Rosenbluth, Rutherford, and Galv{\~{a}}o}}]{Coppi_76}
\bibinfo{author}{\bibfnamefont{B.}~\bibnamefont{Coppi}},
  \bibinfo{author}{\bibfnamefont{R.}~\bibnamefont{Pellat}},
  \bibinfo{author}{\bibfnamefont{M.}~\bibnamefont{Rosenbluth}},
  \bibinfo{author}{\bibfnamefont{P.}~\bibnamefont{Rutherford}},
  \bibnamefont{and}
  \bibinfo{author}{\bibfnamefont{R.}~\bibnamefont{Galv{\~{a}}o}},
  \bibinfo{journal}{Sov. J. Plasma Phys.} \textbf{\bibinfo{volume}{2}},
  \bibinfo{pages}{533} (\bibinfo{year}{1976}).

\bibitem[{\citenamefont{Waelbroeck}(1989)}]{Waelbroeck_89}
\bibinfo{author}{\bibfnamefont{F.}~\bibnamefont{Waelbroeck}},
  \bibinfo{journal}{Phys. Fluids B} \textbf{\bibinfo{volume}{1}},
  \bibinfo{pages}{2372} (\bibinfo{year}{1989}).

\bibitem[{\citenamefont{Militello and Porcelli}(2004)}]{Militello_04}
\bibinfo{author}{\bibfnamefont{F.}~\bibnamefont{Militello}} \bibnamefont{and}
  \bibinfo{author}{\bibfnamefont{F.}~\bibnamefont{Porcelli}},
  \bibinfo{journal}{Phys. Plasmas} \textbf{\bibinfo{volume}{11}},
  \bibinfo{pages}{L13} (\bibinfo{year}{2004}).

\bibitem[{\citenamefont{Escande and Ottaviani}(2004)}]{Escande_04}
\bibinfo{author}{\bibfnamefont{D.}~\bibnamefont{Escande}} \bibnamefont{and}
  \bibinfo{author}{\bibfnamefont{M.}~\bibnamefont{Ottaviani}},
  \bibinfo{journal}{Phys. Lett. A} \textbf{\bibinfo{volume}{323}},
  \bibinfo{pages}{278} (\bibinfo{year}{2004}).

\bibitem[{\citenamefont{Loureiro et~al.}(2005)\citenamefont{Loureiro, Cowley,
  Dorland, Haines, and Schekochihin}}]{LoureiroCowleyDorland_05}
\bibinfo{author}{\bibfnamefont{N.~F.} \bibnamefont{Loureiro}},
  \bibinfo{author}{\bibfnamefont{S.~C.} \bibnamefont{Cowley}},
  \bibinfo{author}{\bibfnamefont{W.}~\bibnamefont{Dorland}},
  \bibinfo{author}{\bibfnamefont{M.~G.} \bibnamefont{Haines}},
  \bibnamefont{and} \bibinfo{author}{\bibfnamefont{A.~A.}
  \bibnamefont{Schekochihin}}, \bibinfo{journal}{Phys. Rev. Lett.}
  \textbf{\bibinfo{volume}{95}}, \bibinfo{pages}{235003}
  (\bibinfo{year}{2005}).

\bibitem[{\citenamefont{Ahedo and Ramos}(2009)}]{AhedoRamos_09}
\bibinfo{author}{\bibfnamefont{E.}~\bibnamefont{Ahedo}} \bibnamefont{and}
  \bibinfo{author}{\bibfnamefont{J.~J.} \bibnamefont{Ramos}},
  \bibinfo{journal}{Plasma Phys. Control. Fusion}
  \textbf{\bibinfo{volume}{51}}, \bibinfo{pages}{055018}
  (\bibinfo{year}{2009}).

\bibitem[{\citenamefont{Fitzpatrick}(2010)}]{Fitzpatrick_10}
\bibinfo{author}{\bibfnamefont{R.}~\bibnamefont{Fitzpatrick}},
  \bibinfo{journal}{Phys. Plasmas} \textbf{\bibinfo{volume}{17}},
  \bibinfo{pages}{042101} (\bibinfo{year}{2010}).

\bibitem[{\citenamefont{Loureiro and Hammett}(2008)}]{LoureiroHammett_08}
\bibinfo{author}{\bibfnamefont{N.~F.} \bibnamefont{Loureiro}} \bibnamefont{and}
  \bibinfo{author}{\bibfnamefont{G.~W.} \bibnamefont{Hammett}},
  \bibinfo{journal}{J. Comp. Phys.} \textbf{\bibinfo{volume}{227}},
  \bibinfo{pages}{4518} (\bibinfo{year}{2008}).

\bibitem[{\citenamefont{{Del Sarto} et~al.}(2011)\citenamefont{{Del Sarto},
  Marchetto, Pegoraro, and Califano}}]{Sarto_11}
\bibinfo{author}{\bibfnamefont{D.}~\bibnamefont{{Del Sarto}}},
  \bibinfo{author}{\bibfnamefont{C.}~\bibnamefont{Marchetto}},
  \bibinfo{author}{\bibfnamefont{F.}~\bibnamefont{Pegoraro}}, \bibnamefont{and}
  \bibinfo{author}{\bibfnamefont{F.}~\bibnamefont{Califano}},
  \bibinfo{journal}{Plasma Phys. Control. Fusion}
  \textbf{\bibinfo{volume}{53}}, \bibinfo{pages}{035008}
  (\bibinfo{year}{2011}).

\bibitem[{\citenamefont{Cowley et~al.}(1986)\citenamefont{Cowley, Kulsrud, and
  Hahm}}]{Cowley_86}
\bibinfo{author}{\bibfnamefont{S.}~\bibnamefont{Cowley}},
  \bibinfo{author}{\bibfnamefont{R.~M.} \bibnamefont{Kulsrud}},
  \bibnamefont{and} \bibinfo{author}{\bibfnamefont{T.~S.} \bibnamefont{Hahm}},
  \bibinfo{journal}{Phys. Fluids B} \textbf{\bibinfo{volume}{29}},
  \bibinfo{pages}{3230} (\bibinfo{year}{1986}).

\bibitem[{\citenamefont{Porcelli}(1991)}]{Porcelli_91}
\bibinfo{author}{\bibfnamefont{F.}~\bibnamefont{Porcelli}},
  \bibinfo{journal}{Phys. Rev. Lett.} \textbf{\bibinfo{volume}{66}},
  \bibinfo{pages}{425} (\bibinfo{year}{1991}).

\bibitem[{\citenamefont{Zakharov and Rogers}(1992)}]{ZakharovRogers_92}
\bibinfo{author}{\bibfnamefont{L.}~\bibnamefont{Zakharov}} \bibnamefont{and}
  \bibinfo{author}{\bibfnamefont{B.}~\bibnamefont{Rogers}},
  \bibinfo{journal}{Phys. Fluids B} \textbf{\bibinfo{volume}{4}},
  \bibinfo{pages}{3285} (\bibinfo{year}{1992}).

\bibitem[{\citenamefont{Zocco and Schekochihin}(2011)}]{ZoccoSchekochihin_11}
\bibinfo{author}{\bibfnamefont{A.}~\bibnamefont{Zocco}} \bibnamefont{and}
  \bibinfo{author}{\bibfnamefont{A.~A.} \bibnamefont{Schekochihin}},
  \bibinfo{journal}{arXiv:1104.4622}  (\bibinfo{year}{2011}).

\bibitem[{\citenamefont{{Numata} et~al.}(2010)\citenamefont{{Numata}, {Howes},
  {Tatsuno}, {Barnes}, and {Dorland}}}]{NumataHowesTatsuno_10}
\bibinfo{author}{\bibfnamefont{R.}~\bibnamefont{{Numata}}},
  \bibinfo{author}{\bibfnamefont{G.~G.} \bibnamefont{{Howes}}},
  \bibinfo{author}{\bibfnamefont{T.}~\bibnamefont{{Tatsuno}}},
  \bibinfo{author}{\bibfnamefont{M.}~\bibnamefont{{Barnes}}}, \bibnamefont{and}
  \bibinfo{author}{\bibfnamefont{W.}~\bibnamefont{{Dorland}}},
  \bibinfo{journal}{J. Comput. Phys.} \textbf{\bibinfo{volume}{229}},
  \bibinfo{pages}{9347} (\bibinfo{year}{2010}).

\bibitem[{\citenamefont{{Abel} et~al.}(2008)\citenamefont{{Abel}, {Barnes},
  {Cowley}, {Dorland}, and {Schekochihin}}}]{AbelBarnesCowley_08}
\bibinfo{author}{\bibfnamefont{I.~G.} \bibnamefont{{Abel}}},
  \bibinfo{author}{\bibfnamefont{M.}~\bibnamefont{{Barnes}}},
  \bibinfo{author}{\bibfnamefont{S.~C.} \bibnamefont{{Cowley}}},
  \bibinfo{author}{\bibfnamefont{W.}~\bibnamefont{{Dorland}}},
  \bibnamefont{and} \bibinfo{author}{\bibfnamefont{A.~A.}
  \bibnamefont{{Schekochihin}}}, \bibinfo{journal}{Phys. Plasmas}
  \textbf{\bibinfo{volume}{15}}, \bibinfo{pages}{122509}
  (\bibinfo{year}{2008}).

\bibitem[{\citenamefont{{Barnes} et~al.}(2009)\citenamefont{{Barnes}, {Abel},
  {Dorland}, {Ernst}, {Hammett}, {Ricci}, {Rogers}, {Schekochihin}, and
  {Tatsuno}}}]{BarnesAbelDorland_09}
\bibinfo{author}{\bibfnamefont{M.}~\bibnamefont{{Barnes}}},
  \bibinfo{author}{\bibfnamefont{I.~G.} \bibnamefont{{Abel}}},
  \bibinfo{author}{\bibfnamefont{W.}~\bibnamefont{{Dorland}}},
  \bibinfo{author}{\bibfnamefont{D.~R.} \bibnamefont{{Ernst}}},
  \bibinfo{author}{\bibfnamefont{G.~W.} \bibnamefont{{Hammett}}},
  \bibinfo{author}{\bibfnamefont{P.}~\bibnamefont{{Ricci}}},
  \bibinfo{author}{\bibfnamefont{B.~N.} \bibnamefont{{Rogers}}},
  \bibinfo{author}{\bibfnamefont{A.~A.} \bibnamefont{{Schekochihin}}},
  \bibnamefont{and}
  \bibinfo{author}{\bibfnamefont{T.}~\bibnamefont{{Tatsuno}}},
  \bibinfo{journal}{Phys. Plasmas} \textbf{\bibinfo{volume}{16}},
  \bibinfo{pages}{072107} (\bibinfo{year}{2009}).

\bibitem[{\citenamefont{Spitzer and H{\"{a}}rm}(1953)}]{SpitzerHarm_53}
\bibinfo{author}{\bibfnamefont{L.}~\bibnamefont{Spitzer}, \bibfnamefont{Jr.}}
  \bibnamefont{and}
  \bibinfo{author}{\bibfnamefont{R.}~\bibnamefont{H{\"{a}}rm}},
  \bibinfo{journal}{Phys. Rev.} \textbf{\bibinfo{volume}{89}},
  \bibinfo{pages}{977} (\bibinfo{year}{1953}).

\bibitem[{\citenamefont{Rogers et~al.}(2007)\citenamefont{Rogers, Kobayashi,
  Ricci, Dorland, Drake, and Tatsuno}}]{RogersKobayashiRicci_07}
\bibinfo{author}{\bibfnamefont{B.~N.} \bibnamefont{Rogers}},
  \bibinfo{author}{\bibfnamefont{S.}~\bibnamefont{Kobayashi}},
  \bibinfo{author}{\bibfnamefont{P.}~\bibnamefont{Ricci}},
  \bibinfo{author}{\bibfnamefont{W.}~\bibnamefont{Dorland}},
  \bibinfo{author}{\bibfnamefont{J.}~\bibnamefont{Drake}}, \bibnamefont{and}
  \bibinfo{author}{\bibfnamefont{T.}~\bibnamefont{Tatsuno}},
  \bibinfo{journal}{Phys. Plasmas} \textbf{\bibinfo{volume}{14}},
  \bibinfo{pages}{092110} (\bibinfo{year}{2007}).

\end{thebibliography}
\end{document}